\documentclass[sigconf]{acmart}

\AtBeginDocument{%
  }

\usepackage{enumitem}
\usepackage{booktabs}
\usepackage{amsmath}
\usepackage{algorithm}
\usepackage{algorithmic}
\usepackage{graphicx}
\usepackage{amsthm}
\usepackage{color}
\usepackage{multicol}
\usepackage{multirow}
\usepackage{arydshln}
\usepackage{subcaption}
\usepackage{colortbl}

\usepackage{balance}

\begin{document}

\title{Taming the Long Tail: Efficient Item-wise Sharpness-Aware Minimization for LLM-based Recommender Systems}


\author{Jiaming Zhang}  
\email{jm.zhang@zju.edu.cn}  
\affiliation{  
  \institution{Zhejiang University}  
  \city{Hangzhou}  
  \state{Zhejiang}  
  \country{China}  
}  

\author{Yuyuan Li}  
\email{y2li@hdu.edu.cn}  
\affiliation{  
  \institution{Hangzhou Dianzi University}  
  \city{Hangzhou}  
  \state{Zhejiang}  
  \country{China}  
}  

\author{Xiaohua Feng}  
\email{fengxiaohua@zju.edu.cn}  
\affiliation{  
  \institution{Zhejiang University}  
  \city{Hangzhou}  
  \state{Zhejiang}  
  \country{China}  
}  

\author{Li Zhang}  
\email{zhanglizl80@gmail.com}  
\affiliation{  
  \institution{Zhejiang University}  
  \city{Hangzhou}  
  \state{Zhejiang}  
  \country{China}  
}  

\author{Longfei Li}  
\email{longyao.llf@antgroup.com}  
\affiliation{  
  \institution{Ant Group}  
  \city{Hangzhou}  
  \state{Zhejiang}  
  \country{China}  
}  

\author{Jun Zhou}  
\email{jun.zhoujun@antgroup.com}  
\affiliation{  
  \institution{Ant Group}  
  \city{Hangzhou}  
  \state{Zhejiang}  
  \country{China}  
}  

\author{Chaochao Chen*}  
\email{zjuccc@zju.edu.cn}  
\affiliation{  
  \institution{Zhejiang University}  
  \city{Hangzhou}  
  \state{Zhejiang}  
  \country{China}  
}  
% 添加通讯作者标识  
\thanks{*Corresponding author}

\renewcommand{\shortauthors}{Jiaming Zhang et al.}

\begin{abstract}
Large Language Model-based Recommender Systems (LRSs) have recently emerged as a new paradigm in sequential recommendation by directly adopting LLMs as backbones. 
While LRSs demonstrate strong knowledge utilization and instruction-following abilities, they have not been systematically studied under the long-standing \textit{long-tail problem}. 
In this paper, we conduct an empirical study and reveal that LRSs face two distinct types of long-tail: i) \textit{prior long-tail}, inherited implicitly from pretraining corpora, and ii) \textit{data long-tail}, originating from skewed recommendation datasets. 
Our analysis shows that both contribute to the performance disparity between head and tail items, with the intersection of the two heads exhibiting an even stronger head effect. Nevertheless, the overall performance distribution in LRSs, especially on the tail, remains dominated by the data long-tail.
To address this challenge, we propose \textit{Efficient Item-wise Sharpness-Aware Minimization (EISAM)}, a novel optimization framework that improves tail-item performance by adaptively regularizing the loss landscape at the item level. 
EISAM introduces an efficient penalty design that captures fine-grained item-specific sharpness while maintaining computational scalability for LLMs. 
In addition, we %provide a theoretical analysis and 
derive a generalization bound for EISAM.
Our theoretical analysis shows that the bound decreases at a faster rate under our item-wise regularization, offering theoretical support for its effectiveness.
Extensive experiments on three real-world datasets demonstrate that EISAM significantly boosts tail-item recommendation performance while preserving overall quality, establishing the first systematic solution to the long-tail problem in LRSs.
\end{abstract}





\maketitle

\section{Introduction}
Recommender Systems (RSs) have been widely deployed in various domains, including news~\cite{turcotte2015news}, videos~\cite{zhou2010impact}, and medications~\cite{bao2016intelligent}. 
Traditional RSs rely heavily on limited interaction data and single-form inputs~\cite{Wang2025Cluster}. They lack broad world knowledge and instruction understanding ability, which limits further improvements in recommendation quality and personalization.  
Large Language Models (LLMs)~\cite{achiam2023gpt,touvron2023llama,luautoannotator}, with strong knowledge and instruction understanding, have recently been integrated into RSs~\cite{xi2024towards,shenglanguage}. Particularly in sequential recommendation, LLMs are directly used as new backbone models, referred to as LLM-based RSs (LRSs)~\cite{bao2023tallrec,bao2025bi}. 

However, this new LRS paradigm has not fully explored long-standing problems in traditional RSs, i.e., the \textit{long-tail problem}.  
Real-world data usually follows a long-tail distribution, while recommendation data is often more skewed, where a very small number of popular items account for the majority of occurrences~\cite{park2008long,yao2024swift}. Thus, in traditional RSs, a small portion of popular items dominates exposure and accuracy, while tail items receive limited attention~\cite{yin2012challenging,jang2020cities,zhao2024contextual}. 
Although LLMs have the theoretical potential to enhance long-tail performance in traditional RSs via the additional auxiliary information~\cite{liu2024llm,wu2024coral,liu2025llmemb,wang2024rcfr}, the long-tail problem in the LRS paradigm has not been explored.  

LRSs are typically obtained by tuning pre-trained LLMs with recommendation data. Consequently, LRSs face two types of long-tail. The first is the long-tail in training data, referred to as \textbf{data long-tail}. The second comes from the prior LLM pretraining corpus, referred to as \textbf{prior long-tail}. %\lyy{are these names already existing? if yes, add ref} 
Since pretraining data are inaccessible or extremely large, and cannot be directly aligned with recommendation data, prior long-tail can only be reflected implicitly through LLM parameters and ultimately manifested in model performance. %\lyy{so what? making long-tail for LRSs more challenging?}
Accordingly, the prior long-tail introduced by pretraining corpus can be inferred from the model’s performance. 

To systematically investigate both types of long-tail effects, we first estimate the prior long-tail based on model performance, and then evaluate the impact of the data long-tail through fine-tuned LRS models. 
\textit{Our empirical study shows that both the prior long-tail and the data long-tail affect the final recommendation performance, yet the data long-tail remains the dominant factor.} 
In both cases, head items consistently outperform tail items, and the intersection of the two heads exhibits the strongest head effect, achieving the best overall performance. However, the data long-tail consistently yields the worst performance, regardless of whether it intersects with the prior head or tail. 
%
% In summary,
These observations demonstrate that the prior distribution mainly contributes to improving head performance, with little additional negative impact on the tail. 
This suggests that, to enhance recommendation quality for tail items, studying and addressing the data long-tail remains crucial. 

Based on this understanding, we propose \textbf{Efficient Item-wise Sharpness-Aware Minimization (EISAM)} to improve long-tail performance in LRSs. Sharpness-Aware Minimization (SAM) is known to enhance generalization by flattening the loss landscape~\cite{foret2021SAM,li2025focal,li2021self}, but its naive adoption in recommendation has two major drawbacks. One is that directly applying SAM to the overall loss~\cite{chen2023sharpness} neglects the distinction between head and tail groups and lacks fine-grained targeted optimization. While this may improve overall performance, it fails to address the inferior performance of tail items. The other is that fine-grained control usually incurs high computational overhead~\cite{wang2024intersectional}, which is particularly problematic for LRSs with massive parameter scales. EISAM addresses both issues by introducing an efficient item-wise sharpness penalty that adaptively regularizes loss curvature per item. This design improves tail-item generalization while maintaining computational efficiency. We further provide a theoretical analysis, showing that EISAM admits a tighter generalization bound and decreases at a faster rate compared to other SAM methods. The implementation and reproducible code are publicly available in our anonymous repository at \url{https://github.com/xiye7lai/samlrs}. 

Our main contributions are summarized as follows:  
\begin{itemize}[leftmargin=*]\setlength{\itemsep}{-\itemsep}
\item To the best of our knowledge, we are the first to systematically investigate the long-tail problem in LRSs, distinguishing between the \textit{prior long-tail} inherited from pretraining and the \textit{data long-tail} in recommendation data.
\item We conduct an empirical study and comprehensive analysis to examine the impacts of both prior and data long-tail effects within the LRS paradigm, revealing that data distribution remains the dominant factor shaping head and tail performance disparities.
\item We introduce Efficient Item-wise Sharpness-Aware Minimization (EISAM), an efficient item-level optimization method that adaptively regularizes loss curvature to enhance tail-item generalization while maintaining computational efficiency.
\item We provide both theoretical and empirical validations, showing that EISAM achieves consistent improvements on multiple real-world datasets, with significant gains on tail items without sacrificing overall recommendation accuracy.
\end{itemize}
\section{Related Work}
\subsection{LLM-based Recommender Systems}
Due to LLMs' powerful capabilities, the utilization of them as a novel backbone for RSs has recently emerged as a promising paradigm. LLM-based RSs directly leverage the generative capability of LLMs to produce recommendations, and this paradigm is particularly suitable for sequential recommendation scenarios since the recommendation process can be naturally formulated in a next-item prediction manner, which aligns well with the intrinsic training objective of LLMs. To adapt LLMs for this purpose, early studies focus on enhancing recommendation generation through the design of effective prompt templates or the incorporation of auxiliary information~\cite{lyu2023llm,shenglanguage,lin2025cec,lu2025dataset}. Mainstream approaches then explore fine-tuning strategies on recommendation datasets~\cite{bao2025bi,zhou2024lidarptq,xu2025avatarshield}, while some works further introduce architectural modifications and special tokens tailored for recommendation tasks~\cite{liao2024llara,qu2024tokenrec}. In addition, preference optimization, which is commonly applied to LLMs after supervised fine-tuning, has also been extended to RSs in both online and offline settings~\cite{rafailov2024direct,meng2024simpo,bai2024aligning,liao2024rosepo}. Beyond aligning LLMs with recommendation objectives, recent efforts further enhance factual accuracy and personalization by integrating techniques such as retrieval-augmented generation~\cite{wang2025knowledge,tong2024mmdfnd}, online feedback mechanisms~\cite{sun2024rlrf4rec,xiao2025points,sun2025yolov4svm}, and multi-agent collaboration~\cite{wang2024macrec,chen2024confusion}.

Despite these advances in boosting recommendation performance, several traditional challenges remain underexplored in this new paradigm, such as the long-tail problem. It is worth noting that while some studies investigate fairness in LRSs~\cite{jiang2024item,gao2025sprec}, their focus is primarily on reducing performance disparities across different groups rather than improving long-tail recommendation quality.

\subsection{Long-tail in Recommender Systems}
Despite the rapid development of RSs, the long-tail phenomenon remains a persistent challenge: a small portion of head items accumulates the majority of interactions, while the vast majority of tail items remain underrepresented, leading to issues in diversity, novelty, and fairness of exposure~\cite{park2008long,anonymous2025comptrack}.

To mitigate this problem, prior research has explored two main directions. The first is knowledge transfer and model design, where information from popular items is propagated to sparse items through shared representations or specialized architectures. Examples include dual-level transfer frameworks linking model- and item-level signals~\cite{zhang2021model,liu2023contrastive,liu2026health}, decoupling-based networks that separately handle head and tail dynamics~\cite{zhang2023empowering,zhang2023multi,su2023enhancing}, and domain transfer methods tailored for sequential recommendation~\cite{wu2022dynamics,chen2022differential,wang2025medical}. The second is debiasing through causal inference, which disentangles genuine user preference from popularity-driven exposure via reweighting, counterfactual estimation, or deconfounded objectives~\cite{xia2023user,liu2024interact,zhang2025yoloppa}.

However, these approaches are largely developed under conventional RS backbones, making them difficult to adapt in the emerging LLM-based paradigm. Techniques such as architectural decoupling or causal reweighting require direct control over model structures, training objectives, or interaction-level sampling, whereas LRSs typically operate in a generative manner with frozen backbones or lightweight tuning. This mismatch limits the applicability of existing long-tail solutions. While some recent studies have attempted to integrate LLMs into traditional RSs through enhanced representations or retrieval-augmented generation~\cite{wu2024coral,liu2024llm,liu2025llmemb,li2024instant3d}, there has been little to no exploration of how long-tail challenges can be systematically addressed within LRSs.
\section{Problem Setup}
In LRSs, the backbone relies solely on item information, and the recommendation task is formulated as sequential next-item prediction. Since user features are not explicitly modeled, we focus exclusively on the \textit{item long-tail problem}.  

Let the item space be \(\mathcal{I} = \{i_1, i_2, \dots, i_{|\mathcal{I}|}\}\). Each recommendation instance is an item sequence \(s = \{i_1, i_2, \dots, i_{L}\}\), where \(L\) denotes the sequence length. Given a prefix sequence \(s\), the task is to predict the next item from the item space:  
\begin{equation}
\hat{i} = \arg\max_{i \in \mathcal{I}} P(i \mid s).
\end{equation}
We denote the training dataset as \(\mathcal{S} = \{(s_j, i_j)\}_{j=1}^N\), where \(s_j\) is an input sequence and \(i_j\) is the corresponding target item. The training loss is denoted as \(\ell(\boldsymbol{\theta}; s, i)\), which in LRSs typically corresponds to the supervised fine-tuning (SFT) loss with model parameters \(\boldsymbol{\theta}\). The empirical training loss over dataset \(\mathcal{S}\) is then defined as  
\[
L_{\mathcal{S}}(\boldsymbol{\theta}) \triangleq \frac{1}{N} \sum_{j=1}^N \ell(\boldsymbol{\theta}; s_j, i_j).
\]  

In practice, the distribution of target items is highly imbalanced. For each item \(i \in \mathcal{I}\), let \(n_i\) denote the number of times \(i\) appears as a prediction target in the training set, and define its empirical frequency as \(q_i = n_i / N\), where \(N = \sum_{i \in \mathcal{I}} n_i\). Ordering items such that \(n_1 \geq n_2 \geq \dots \geq n_{|\mathcal{I}|}\), the empirical distribution \(\mathcal{D}\) reveals a long-tailed pattern in which the head items occupy the majority.  

Following prior work in long-tailed recognition, we also define an \textit{ideal balanced distribution} \(\mathcal{D}_{bal}\), under which all items occur with equal probability. The expected population loss under \(\mathcal{D}\) and \(\mathcal{D}_{bal}\) can be written as  
\[
L_{\mathcal{D}}(\boldsymbol{\theta}) = \mathbb{E}_{(s,i) \sim \mathcal{D}} \big[\ell(\boldsymbol{\theta}; s,i)\big], \quad
L_{\mathcal{D}_{bal}}(\boldsymbol{\theta}) = \mathbb{E}_{(s,i) \sim \mathcal{D}_{bal}} \big[\ell(\boldsymbol{\theta}; s,i)\big].
\]  

The long-tail challenge in LRSs is therefore to design learning strategies that reduce the gap between \(L_{\mathcal{D}}(\boldsymbol{\theta})\) and \(L_{\mathcal{D}_{bal}}(\boldsymbol{\theta})\), so that the RS maintains accuracy on head items while also improving performance on tail items that are severely underrepresented.
\section{Empirical Study}

To investigate the long-tail phenomenon in LRSs, we conduct an empirical analysis using the widely adopted BIGRec~\cite{bao2025bi} model. Details can be found in Section~\ref{sec:exp}.  

\begin{figure}[t]
  \centering

  \begin{subfigure}[t]{0.58\linewidth}
    \centering
    \includegraphics[width=\linewidth]{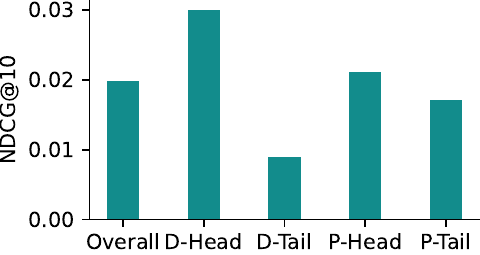}
  \end{subfigure}\hfill
  \begin{subfigure}[t]{0.4\linewidth}
    \centering
    \includegraphics[width=\linewidth]{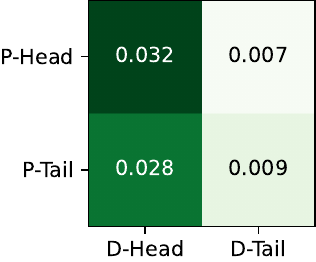}
  \end{subfigure}

  \caption{Performance across data/prior groups.}
  \label{fig:prior}
\end{figure}

To examine the \textit{prior long-tail}, we first utilize the base LLM to directly generate recommendation results. However, since the raw generation is not aligned with the sequential recommendation input format and lacks an explicit item list for reference, the generated outputs exhibit very low accuracy and extremely high redundancy. To address this issue, we first design prompts that explicitly instruct the base LLM to generate $k$ distinct recommendation items. To further align the outputs with the sequential recommendation format, we perform a lightweight fine-tuning using a small number of synthetic examples that directly provide $k$ non-redundant recommendations in the required format. Following the grounding strategy proposed in BIGRec, the generated $k$ outputs are then mapped to the closest items in the candidate set by comparing their representations with item embeddings obtained from feeding items into the LLM, and the most similar items are selected as the predicted targets. By aggregating these predicted targets across all sequences in the dataset, we obtain the prior long-tail distribution. We report in Figure~\ref{fig:prior} the performance (evaluated by NDCG@10) of LRS fine-tuned with recommendation data under both prior long-tail and data long-tail distributions.

From the empirical results, we observe that both the prior long-tail (P) and the data long-tail (D) influence LRS performance, yet the data long-tail remains the dominant factor. The best performance occurs in the head--head intersection ($P$-Head $\cap$ $D$-Head), indicating that these items benefit simultaneously from training bias and the LLM's inherent tendency to answer head items correctly. However, the lowest accuracy does not appear in the tail--tail intersection, but rather in regions involving data-tail items, especially $P$-Head $\cap$ $D$-Tail. This shows that once items fall into the data-tail space, the model's ability to recommend them drops sharply, regardless of their prior classification. The prior head category also contains some low-activity items that are actually data-tail in nature, reflecting that the LLM's prior knowledge is only partially aligned with the recommendation domain. When intersecting with the data distribution, these items are still significantly dominated by data-tail effects, and because they participate in training much less frequently, their recommendation quality improves far more slowly than that of true head items.

\textbf{In summary,} we summarize our main findings as follows.
\begin{itemize}[leftmargin=*]\setlength{\itemsep}{-\itemsep}
\item Within the LRS paradigm, the performance gap between head and tail items is primarily determined by the data distribution.
\item The prior head items, especially when interacting with data head items can further amplify head performance.
\item The prior long-tail has only a minimal negative effect on tail items during training.
\end{itemize}

Therefore, to mitigate long-tail issues in LRSs, we should treat the data long-tail distribution as the primary target and improve the representation and performance of data-tail items.

\section{Methodology}
We propose Efficient Item-wise Sharpness-Aware Minimization (EISAM) to mitigate the item long-tail problem in LRSs. 
The key idea is to regularize the loss landscape at the item level in a frequency-aware manner, while retaining training efficiency. 
Our method consists of two parts: \textit{item-wise sharpness regularization} and an \textit{efficient optimization procedure}.

\subsection{Item-wise Sharpness Regularization}
For each item $i\in\mathcal{I}$, denote by $\mathcal{S}(i)=\{(s,i)\in\mathcal{S}\}$ the subset where $i$ is the target, and define its item-wise empirical loss
\begin{equation}
L_{\mathcal{S}}^{(i)}(\boldsymbol{\theta}) \triangleq \frac{1}{|\mathcal{S}(i)|}\sum_{(s,i)\in\mathcal{S}(i)} \ell(\boldsymbol{\theta}; s,i).
\end{equation}

We introduce \textit{item-wise sharpness} to measure how sharp the loss landscape of item $i$ is around the current parameters. 
It is defined as the loss difference between the original LRS parameters and their perturbed counterpart.
Given a perturbation $\epsilon$ with $\|\epsilon\|\le \rho$, the item-wise sharpness is
\begin{equation}
    IS^{(i)}(\boldsymbol{\theta},\epsilon)
    \triangleq L_{\mathcal{S}}^{(i)}(\boldsymbol{\theta}+\epsilon) - L_{\mathcal{S}}^{(i)}(\boldsymbol{\theta}).
\end{equation}

To emphasize tail items, we introduce a frequency-dependent weighting function $f(\cdot)$ that assigns larger weights to less frequent items. 
Specifically, for an item with empirical frequency $q_i$, we use a monotonically decreasing function $f(q_i)$ (e.g., $f(q_i)=q_i^{-\alpha}$ or $f(q_i)=-\log q_i$).
By combining all item-wise sharpness terms with their weights and aggregating them inside the maximization, we obtain the overall item-wise sharpness regularization:
\begin{equation}
L^{\text{SAM}}_{\mathcal{S}}(\boldsymbol{\theta})
= \max_{\|\epsilon\|\le \rho}  
\sum_{i\in\mathcal{I}} f(q_i) IS^{(i)}(\boldsymbol{\theta},\epsilon).
\end{equation}

EISAM then integrates the empirical risk with this regularization:
\begin{equation}
J(\boldsymbol{\theta})  =   L_{\mathcal{S}}(\boldsymbol{\theta})  +  \lambda L^{\text{SAM}}_{\mathcal{S}}(\boldsymbol{\theta}),
\end{equation}
where $\lambda$ is a hyperparameter that controls the strength of the item-wise sharpness regularization. The weighting function $f(\cdot)$  provides fine-grained control. By selecting different functional forms or adjusting its hyperparameters, we can tune how strongly the regularization emphasizes tail items depending on the severity of the long-tail distribution. Item-wise sharpness regularization improves the flatness of the loss landscape for tail items, thereby enhancing recommendation performance on the long tail while maintaining the performance on head items. 

\subsection{Efficient Optimization Procedure}
The definition of $L^{\text{SAM}}_{\mathcal{S}}(\boldsymbol{\theta})$ involves an inner maximization over the perturbation $\epsilon$, which cannot be solved exactly in practice. 
To make the problem tractable, we approximate the change in each item-wise loss under perturbation by its first-order Taylor expansion around $\boldsymbol{\theta}$:
\begin{equation}
L_{\mathcal{S}}^{(i)}(\boldsymbol{\theta}+\epsilon) \approx L_{\mathcal{S}}^{(i)}(\boldsymbol{\theta}) + \epsilon^\top \nabla_{\boldsymbol{\theta}} L_{\mathcal{S}}^{(i)}(\boldsymbol{\theta}).
\end{equation}
Substituting this into the definition of $L^{\text{SAM}}_{\mathcal{S}}(\boldsymbol{\theta})$ yields
\begin{equation}
L^{\text{SAM}}_{\mathcal{S}}(\boldsymbol{\theta})   \approx  
\max_{\|\epsilon\|\le \rho}  
\epsilon^\top \Bigg(\sum_{i\in\mathcal{I}} f(q_i) \nabla_{\boldsymbol{\theta}} L_{\mathcal{S}}^{(i)}(\boldsymbol{\theta})\Bigg).
\end{equation}
This expression shows that the worst-case perturbation direction depends only on the weighted sum of item-wise gradients. 
Under the $\ell_2$ constraint, the closed-form solution is
\begin{equation}
\hat{\epsilon}(\boldsymbol{\theta}) = \rho  
\frac{\sum_{i\in\mathcal{I}} f(q_i) \nabla_{\boldsymbol{\theta}} L_{\mathcal{S}}^{(i)}(\boldsymbol{\theta})}
{\left\|\sum_{i\in\mathcal{I}} f(q_i) \nabla_{\boldsymbol{\theta}} L_{\mathcal{S}}^{(i)}(\boldsymbol{\theta})\right\|_2}.
\end{equation}
Intuitively, $\hat{\epsilon}(\boldsymbol{\theta})$ points in the ascent direction that causes the largest increase in the weighted loss across both head and tail items.  

We can now compute the gradient of $J(\boldsymbol{\theta})$. However, computing $\nabla_{\boldsymbol{\theta}} L^{\text{SAM}}_{\mathcal S}(\boldsymbol{\theta})$ directly poses a difficulty, since the perturbation $\hat\epsilon(\boldsymbol{\theta})$ itself depends on the weighted item-wise loss, which for simplicity we denote as $L^w_{\mathcal{S}}(\boldsymbol{\theta})=\sum_{i\in\mathcal{I}} f(q_i)L_{\mathcal S}^{(i)}(\boldsymbol{\theta})$. Differentiating $L^{\text{SAM}}_{\mathcal S}(\boldsymbol{\theta})$ under this dependence leads to
\begin{equation}
\nabla_{\boldsymbol{\theta}} L^{\text{SAM}}_{\mathcal S}(\boldsymbol{\theta}) =
\Big(I+\tfrac{\partial \hat\epsilon}{\partial \boldsymbol{\theta}}\Big)^\top
\nabla_{\boldsymbol{\theta}} L^w_{\mathcal S}(\boldsymbol{\theta}+\hat\epsilon(\boldsymbol{\theta}))
-
\nabla_{\boldsymbol{\theta}} L^w_{\mathcal S}(\boldsymbol{\theta}),
\end{equation}
where $I$ denotes the identity matrix. Computing the above gradient would involve Hessian operations and thus be computationally expensive. As in most SAM-based methods, we drop these higher-order terms and treat $\hat\epsilon(\boldsymbol{\theta})$ as constant during backpropagation. This yields the practical approximation
\begin{equation}
\nabla_{\boldsymbol{\theta}} L^{\text{SAM}}_{\mathcal S}(\boldsymbol{\theta}) 
  \approx   
\left.\nabla_{\boldsymbol{\theta}} L^w_{\mathcal S}(\boldsymbol{\theta})\right|_{\boldsymbol{\theta}+\hat\epsilon(\boldsymbol{\theta})}
  -  
\nabla_{\boldsymbol{\theta}} L^w_{\mathcal S}(\boldsymbol{\theta})|_{\boldsymbol{\theta}}.
\end{equation}

Substituting into the definition of $J(\boldsymbol{\theta})$, the full gradient becomes
\begin{equation}
\nabla_{\boldsymbol{\theta}} J(\boldsymbol{\theta}) \approx 
\left . \big[\nabla_{\boldsymbol{\theta}} L_{\mathcal S}(\boldsymbol{\theta}) - \lambda \nabla_{\boldsymbol{\theta}} L^w_{\mathcal S}(\boldsymbol{\theta})\big]\right|_{\boldsymbol{\theta}}
+
\lambda \left.\nabla_{\boldsymbol{\theta}} L^w_{\mathcal S}(\boldsymbol{\theta})\right|_{\boldsymbol{\theta}+\hat\epsilon(\boldsymbol{\theta})}.
\end{equation}
In practice, solving the EISAM objective therefore requires three backpropagations per iteration:  
\begin{itemize}[leftmargin=*]\setlength{\itemsep}{-\itemsep}
    \item Perturbation construction: compute $\nabla_{\boldsymbol{\theta}} L^w_{\mathcal S}(\boldsymbol{\theta})$ to form $\hat\epsilon(\boldsymbol{\theta})$.  
    \item Unperturbed difference gradient: compute $\nabla_{\boldsymbol{\theta}} L_{\mathcal S}(\boldsymbol{\theta})-\lambda \nabla_{\boldsymbol{\theta}} L^w_{\mathcal S}(\boldsymbol{\theta})$ at $\boldsymbol{\theta}$.  
    \item Perturbed SAM gradient: compute $\nabla_{\boldsymbol{\theta}} L^w_{\mathcal S}(\boldsymbol{\theta})$ at $\boldsymbol{\theta}+\hat\epsilon(\boldsymbol{\theta})$.  
\end{itemize}
Together, these steps provide the full gradient needed to update the parameters in $J(\boldsymbol{\theta})$, thus efficiently solving the EISAM optimization problem. Thus, while EISAM requires slightly more computation than standard training, it remains efficient and scalable, and it explicitly incorporates item-level sharpness information into optimization. The entire process is summarized in Algorithm \ref{alg:eisam}. 
\begin{algorithm}[t]
\caption{EISAM}
\label{alg:eisam}
\raggedright
\textbf{Input:} Training set $\mathcal{S}$, perturbation radius $\rho$, hyperparameter $\lambda$, weighting $f(\cdot)$, learning rate $\eta$, total steps $T$. \\
\textbf{Output:} Optimized Model Parameters $\boldsymbol{\theta}_T$.

\begin{algorithmic}[1]
\STATE Initialize model parameters $\boldsymbol{\theta}_0$;
\FOR{$t$ in $1 \dots T$}
    \STATE Sample batch $\mathcal{B}=\{(s_k,i_k)\}_{k=1}^{B}$;
    \STATE Compute $L_{\mathcal{B}}(\boldsymbol{\theta})$ and $L_{\mathcal{B}}^{w}(\boldsymbol{\theta})$;
    \STATE Compute $\nabla_{\boldsymbol{\theta}} L_{\mathcal{B}}^{w}(\boldsymbol{\theta})$ and
           $\hat{\epsilon}(\boldsymbol{\theta})=\rho \frac{\nabla_{\boldsymbol{\theta}} L_{\mathcal{B}}^{w}(\boldsymbol{\theta})}{\|\nabla_{\boldsymbol{\theta}} L_{\mathcal{B}}^{w}(\boldsymbol{\theta})\|_2}$;
    \STATE Perturb $\boldsymbol{\theta}$ with $\hat{\epsilon}(\boldsymbol{\theta})$ and compute
           $g_1=\left.\nabla_{\boldsymbol{\theta}} L_{\mathcal{B}}^{w}(\boldsymbol{\theta})\right|_{\boldsymbol{\theta}+\hat{\epsilon}(\boldsymbol{\theta})}$;
    \STATE Compute $g_2=\left.\nabla_{\boldsymbol{\theta}}\!\big[L_{\mathcal{B}}(\boldsymbol{\theta})-\lambda L_{\mathcal{B}}^{w}(\boldsymbol{\theta})\big]\right|_{\boldsymbol{\theta}}$;
    \STATE Update parameters: $\boldsymbol{\theta}_{t+1}=\boldsymbol{\theta}_t-\eta[\lambda g_1+g_2]$;
\ENDFOR
\STATE \textbf{return} Optimized Model $\boldsymbol{\theta}_T$
\end{algorithmic}
\end{algorithm}

\subsection{Theoretical Analysis}
We provide theoretical guarantees for EISAM. Our analysis shows that the proposed item-wise sharpness regularization improves flatness on tail items and leads to a tighter generalization bound under the balanced test distribution. The detailed proof can be found in the Appendix.

\paragraph{Notation.}
We assume $\ell(\boldsymbol{\theta};s,i)\in[0,B]$. 
Write $B^w=\sum_{i\in\mathcal{I}} f(q_i) q_i B$. 
Let $H^w(\boldsymbol{\theta})$ denote the Hessian of $L^w_\mathcal{D}(\boldsymbol{\theta})$ at $\boldsymbol{\theta}$, 
$d$ the parameter dimension. We define 
\[
\Phi_D^\lambda(\boldsymbol{\theta})=L_\mathcal{D}(\boldsymbol{\theta})-\lambda L^w_\mathcal{D}(\boldsymbol{\theta}), 
\quad 
\Phi_{\mathcal S}^\lambda(\boldsymbol{\theta})=L_{\mathcal S}(\boldsymbol{\theta})-\lambda L^w_{\mathcal S}(\boldsymbol{\theta}).
\]

We now state three key lemmas that form the basis of our main theorem. The first lemma shows that the adjusted risk concentrates around its empirical version.

\begin{lemma}[Concentration of weighted risks]
\label{lem:concentration}
Following Bernstein’s Inequality~\cite{bernstein1924modification}, with probability $1-\delta$ over the choice of the training set $\mathcal{S} \sim \mathcal{D}$
\begin{equation}
\Phi_\mathcal{D}^\lambda(\boldsymbol{\theta})\ \le\ 
2 \Phi_{\mathcal S}^\lambda(\boldsymbol{\theta})\ +\ 
\frac{40 (B+\lambda B^w)}{3n} \log\frac{2}{\delta}.
\end{equation}
\end{lemma}

Lemma~\ref{lem:concentration} controls the deviation between the population adjusted risk and its empirical counterpart. 
% To convert population perturbed risks into an empirical, radius-bounded training objective, which is exactly the item-wise sharpness part used by EISAM, we next upper bound the expected weighted risk under Gaussian perturbations by the empirical worst-case risk within an $\ell_2$ ball.

\begin{lemma}[Empirical control of weighted perturbed risks]
\label{lem:perturbed-empirical}
Following SAM~\cite{foret2021SAM}, fix any radius $\rho>0$ and define $\sigma_Q=\rho/(\sqrt d+\sqrt{2\ln n})$. 
With probability at least $1-\delta$ over the choice of the training set $\mathcal S\sim\mathcal D$, the following holds:
\begin{equation}
\mathbb{E}_{\epsilon\sim\mathcal N(0,\sigma_Q^2 I)}\Big[L^w_{\mathcal D}(\boldsymbol{\theta}+\epsilon)\Big]
\ \le\
\max_{\|\epsilon\|_2\le \rho}  2 L^w_{\mathcal S}(\boldsymbol{\theta}+\epsilon)
\ +\ \frac{\mathcal C(\boldsymbol{\theta},\rho,d,n,\delta)}{n},
\end{equation}
where
\begin{align}
\mathcal C(\boldsymbol{\theta},\rho,d,n,\delta)
 = &
2+2B^w
 + 2d\ln\!\Big(1+\frac{\|\boldsymbol{\theta}\|_2^2}{d\rho^2}\Big)
 + 4d\log\!\Big(\sqrt d+\sqrt{2\ln n}\Big) \nonumber\\
& + 4\log\!\Bigg(\frac{\pi^2 \sqrt n \big(1+n B^w\big)^2}{3\delta}\Bigg).
\end{align}
\end{lemma}

Lemma~\ref{lem:perturbed-empirical} still bounds a perturbed population quantity. 
% To translate this bound back to the unperturbed population weighted risk and make explicit how curvature improves the bound, we use a second-order expansion that introduces the trace of the Hessian and yields a curvature-dependent bonus term.

\begin{lemma}[Second-order de-noising via curvature]
\label{lem:taylor-curvature}
Following~\cite{li2025focal}, fix any radius $\rho>0$ and let $\sigma_Q=\rho/(\sqrt d+\sqrt{2\ln n})$. 
There exists a remainder term $o\!\left(\tfrac{d\rho^2}{(\sqrt d+\sqrt{2\ln n})^2}\right)$ such that, with probability at least $1-\delta$ over the draw of $\mathcal S\sim\mathcal D$,
\begin{align}
L^w_{\mathcal D}(\boldsymbol{\theta})
\ \le\ &
\ \max_{\|\epsilon\|_2\le \rho}\, 2 L^w_{\mathcal S}(\boldsymbol{\theta}+\epsilon) -\ \frac{\rho^2}{2\big(\sqrt d+\sqrt{2\ln n}\big)^2}\,\mathrm{tr}\!\big(H^w(\boldsymbol{\theta})\big)
\ \nonumber\\
&\  +\ \frac{\mathcal C(\boldsymbol{\theta},\rho,d,n,\delta)}{n}
\ -\ o\!\Big(\tfrac{d\rho^2}{(\sqrt d+\sqrt{2\ln n})^2}\Big).
\end{align}
where the complexity term $\mathcal C(\boldsymbol{\theta},\rho,d,n,\delta)$ is the same as in Lemma~\ref{lem:perturbed-empirical}.
\end{lemma}

We now combine Lemma~\ref{lem:concentration} with Lemma~\ref{lem:taylor-curvature}. Finally, we relate the training distribution to the balanced test distribution using the inequality $L_{\mathcal D}(\boldsymbol{\theta})\ge |\mathcal I| q_{\min} L_{\mathcal D_{\mathrm{bal}}}(\boldsymbol{\theta})$.

\begin{theorem}[Generalization bound of EISAM]
\label{thm:eisam}
Inspired by~\cite{wang2023unified}, let $q_{\min}=\min_{i\in\mathcal I} q_i$. 
Assume the balanced test distribution places equal probability on each item. 
Then for any $\lambda\ge 0$ and any radius $\rho>0$, with probability at least $1-\delta$ over the draw of $\mathcal S\sim\mathcal D$,
\begin{align}
L_{\mathcal D_{\mathrm{bal}}}(\boldsymbol{\theta})
\ \le\ 
&\ \frac{2}{|\mathcal I| q_{\min}}J_\mathcal{S}(\boldsymbol{\theta})  -\ \frac{\lambda \rho^2}{2 |\mathcal I| q_{\min} \big(\sqrt d+\sqrt{2\ln n}\big)^2} \mathrm{tr}\!\big(H^w(\boldsymbol{\theta})\big) \nonumber \\
&\ +\ \frac{1}{|\mathcal I| q_{\min}}\Bigg[
\frac{40 (B+\lambda B^w)}{3n}\ln\frac{2}{\delta} + \frac{\lambda \mathcal C(\boldsymbol{\theta},\rho,d,n,\delta)}{n}
\Bigg] \nonumber\\
&\ -\ \frac{\lambda}{|\mathcal I| q_{\min}} o\!\Big(\tfrac{d\rho^2}{(\sqrt d+\sqrt{2\ln n})^2}\Big).
\end{align}
Here 
$
J_{\mathcal S}(\boldsymbol{\theta}) = L_{\mathcal{S}}(\boldsymbol{\theta}) + \lambda L^{\mathrm{SAM}}_{\mathcal S}(\boldsymbol{\theta}).
$
\end{theorem}

\noindent\textbf{Remarks.}
\begin{itemize}[leftmargin=*]\setlength{\itemsep}{-\itemsep}
% \item \textit{Curvature improves the bound.} The negative term proportional to $\mathrm{tr}(H^w(\boldsymbol{\theta}))$ shows that flatter minima (smaller trace of the Hessian under item-wise weighting) lead to a tighter bound under the balanced test distribution.
\item \textit{Complexity.} All distribution-dependent corrections scale as $O(1/n)$, through the concentration term in Lemma~\ref{lem:concentration} and the complexity $\mathcal C(\boldsymbol{\theta},\rho,d,n,\delta)$ in Lemma~\ref{lem:perturbed-empirical}, making the bound sharper with more data.
\item \textit{Effect of item weights.} The quantity $B^w=\sum_{i\in\mathcal I} f(q_i) q_i B$ captures how the weighting emphasizes tail items. Choosing $f(\cdot)$ that increases the relative weight of tail items reduces imbalance-induced slack and tightens the balanced-distribution bound.
\item \textit{Choice of $\lambda$ and $\rho$.} Larger $\rho$ and $\lambda$ strengthen the curvature bonus but also enlarge the complexity terms; in practice they should be tuned to balance empirical fit and regularization.
\end{itemize}

\section{Experiments}
\label{sec:exp}
We conduct comprehensive experiments on the long-tailed recommendation scenario to evaluate the effectiveness of EISAM.
% In particular, we are interested in examining whether the proposed item-wise sharpness regularization can better handle tail items. 
Specifically, we aim to answer the following research questions:

\begin{itemize}[leftmargin=*]\setlength{\itemsep}{-\itemsep}
    \item \textbf{RQ1.} How does the overall performance of EISAM and its performance on tail items when applied to popular LRS backbones?
    \item \textbf{RQ2.} How does the proposed item-wise sharpness regularization influence the flatness of tail items in LRS?
    \item \textbf{RQ3.} What is the computational efficiency of EISAM?
    \item \textbf{RQ4.} How sensitive is EISAM to different hyperparameter choices under the long-tailed setting?
\end{itemize}
\subsection{Experimental Settings}
\subsubsection{Datastes}
\begin{table}
\centering
\caption{Summary of datasets.}
\label{tab:dataset}
\begin{tabular}{lccc}  
\toprule
Dataset   & \# Items    & \# Interactions & \# Sequences \\
\midrule
ML-1M  & 3,883 & 1,000,209  & 939,809  \\
Steam  &  32,135 & 1,307,310   & 138,970 \\
ADM  &  266,414 & 836,006   & 22,982 \\
\bottomrule
\end{tabular}
\end{table}
We conduct experiments on three widely used real-world datasets in recommendation research.  
\textbf{ML-1M}\footnote{https://grouplens.org/datasets/movielens/}: MovieLens-1M is a benchmark dataset containing one million ratings.  
\textbf{Steam}\footnote{https://cseweb.ucsd.edu/~jmcauley/datasets.html}: The Steam dataset contains reviews and play histories from the Steam video game platform.  
\textbf{ADM}\footnote{http://jmcauley.ucsd.edu/data/amazon/}: The Amazon Digital Music (ADM) dataset includes ratings for digital music items.  

The statistics of these datasets are summarized in Table~\ref{tab:dataset}. To simulate sequential recommendation scenarios, we construct interaction sequences based on timestamps and split them into training, validation, and testing sets with a ratio of 8:1:1. Following prior work on long-tail recommendation, we remove items with fewer than five interaction records to alleviate sparsity. We set the maximum sequence length to 10 across all datasets. Based on the Pareto principle~\cite{box1986analysis}, we designate the top 20\% most frequent items as head items and the remaining 80\% as tail items.  

\subsubsection{LRS Models}
We select two representative LRS backbones:  
\textbf{BIGRec}~\cite{bao2025bi} adopts a two-stage grounding strategy: it first applies supervised fine-tuning (SFT) to align the LLM with the recommendation domain, and then grounds into the item domain by generating embeddings and matching model outputs with item representations.  
\textbf{TALLRec}~\cite{bao2023tallrec} extends LLMs through lightweight fine-tuning combined with a dual-stage paradigm that incorporates instruction-following and recommendation-specific tuning. In our implementation, we further employ TALLRec to generate multiple items simultaneously and adopt beam search to enhance the diversity and quality of the recommended results.
% Following the settings of BIGRec and TALLRec, we first perform instruction fine-tuning and then train the models in an SFT manner on 65,536 sampled sequences without altering the overall data distribution (for ADM, we use all sequences since the dataset contains fewer than 65,536).  

For the base LLM, we adopt the representative open-source models: Llama2-7B (Llama)~\cite{touvron2023llama}.  

\subsubsection{Baselines}
We compare our proposed approach against the following baselines:  
\begin{itemize}[leftmargin=*]\setlength{\itemsep}{2pt}
    \item \textbf{RW}: Re-weighting~\cite{jiang2024item} improves long-tail performance by adjusting training weights according to group popularity.  
    \item \textbf{SAM}: Sharpness-Aware Minimization~\cite{foret2021SAM} is originally introduced to improve the performance of low-degree nodes in graph-based collaborative filtering~\cite{chen2023sharpness}. We extend it to LRSs.  
    \item \textbf{Group SAM}: Group-wise Sharpness-Aware Minimization~\cite{wang2024intersectional} is originally proposed to discover the performance of disadvantaged groups. We adapt it to the LRS paradigm.  
\end{itemize}

\subsubsection{Evaluation Metrics}
To evaluate recommendation performance, we adopt two widely used metrics: Normalized Discounted Cumulative Gain ($\mathrm{NDCG}@K$) and Hit Ratio ($\mathrm{HR}@K$). We set $K=10$.

\subsection{Implement Details}
\begin{figure}[t]
  \centering

  \begin{subfigure}[t]{0.3\linewidth}
    \centering
    \includegraphics[width=\linewidth]{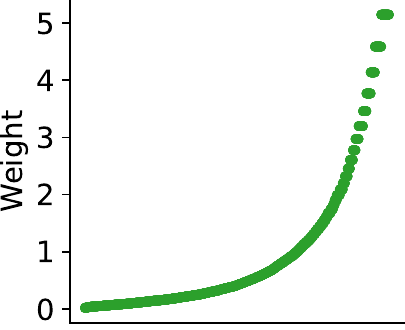}
    \caption{Normalized}
  \end{subfigure}
  \begin{subfigure}[t]{0.3\linewidth}
    \centering
    \includegraphics[width=\linewidth]{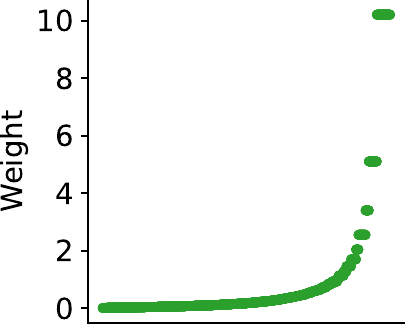}
    \caption{Effective number}
  \end{subfigure}
  \begin{subfigure}[t]{0.3\linewidth}
    \centering
    \includegraphics[width=\linewidth]{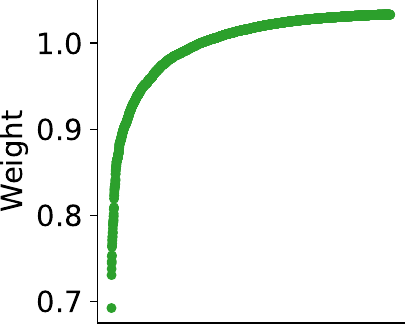}
    \caption{Exponential}
  \end{subfigure}

  \caption{Weight distributions of different weighting functions on the ML-1M dataset, where the x-axis denotes item indices sorted by descending frequency.}
  \label{fig:choose_weight}
\end{figure}
Following prior work~\cite{bao2025bi}, we adopt consistent training configurations. 
For ML-1M and Steam, the number of training epochs is set to 3, while for ADM we train for 2 epochs. 
We use the Adam optimizer with a learning rate of $5\times10^{-4}$ and a batch size of 64. 
All models are fine-tuned with LoRA, where the rank $r$ is set to 8 and the LoRA scaling factor $\alpha$ is 16. 

For the choice of the weighting function $f(\cdot)$, we consider three representative forms: 
(\textit{i}) a variant based solely on normalized sample frequency~\cite{jiang2024item,li2021exploiting}, 
which directly assigns smaller weights to frequent items and larger weights to rare ones, 
formulated as $f_{\text{norm}}(q_i) = \frac{1}{q_i + \epsilon}$, where $q_i$ denotes the empirical frequency of item $i$ and $\epsilon$ is a small constant to ensure numerical stability.

(\textit{ii}) the effective number formulation~\cite{cui2019class}, 
which computes the effective contribution of each item by considering its diminishing information gain with repeated occurrences, 
defined as $f_{\text{eff}}(q_i) = \frac{1 - \beta}{1 - \beta^{q_i}}$, where $\beta \in [0,1)$ controls the smoothness of frequency attenuation.

(\textit{iii}) a smoothly attenuated exponential form~\cite{lin2017focal}, 
which introduces a soft suppression factor that gradually reduces the influence of high-frequency items, 
expressed as $f_{\text{exp}}(q_i) = (1 - q_i)^{\gamma}$
where $\gamma > 0$, adjusts the degree of suppression.

These three weighting schemes produce distinct frequency–weight mappings, as illustrated in Figure~\ref{fig:choose_weight}.
Since sequential recommendation inherently exhibits a long-tailed distribution and each item occurs only a limited number of times, 
overly aggressive weighting strategies tend to make the model focus excessively on a few noisy tail items while ignoring the head items that contribute substantially to overall performance. 
Therefore, we adopt the exponential weighting function as a balanced and stable choice.

\begin{table*}[t]
\centering
\caption{
Performance comparison across datasets and base LRSs. 
The best results are highlighted in bold, and the second-best results are underlined. 
For the Overall and Tail metrics, we additionally report the relative improvements (\%) of EISAM over the best baseline.
}
% \resizebox{\textwidth}{!}{%
\begin{tabular}{ll *{6}{c}} \toprule \multicolumn{2}{c}{} & \multicolumn{2}{c}{Overall} & \multicolumn{2}{c}{Tail} & \multicolumn{2}{c}{Head} \\ \cmidrule(lr){3-4}\cmidrule(lr){5-6}\cmidrule(lr){7-8} \textbf{Dataset} & \textbf{Method} & NDCG@10 & HR@10 & NDCG@10 & HR@10 & NDCG@10 & HR@10 \\ \midrule \multirow{10}{*}{\textbf{ML-1M}} & TALLRec & 0.0197 & 0.0239 & 0.0071 & 0.0101 & \textbf{0.0318} & \textbf{0.0371} \\ & +RW & 0.0149 & 0.0177 & 0.0087 & 0.0111 & 0.0208 & 0.0240 \\ & +SAM & \underline{0.0202} & \underline{0.0241} & 0.0084 & 0.0108 & 0.0315 & 0.0368 \\ & +GroupSAM & 0.0195 & 0.0235 & \underline{0.0090} & \underline{0.0129} & 0.0295 & 0.0336 \\ & \cellcolor{green!10}+EISAM & \cellcolor{green!10}\textbf{0.0211(+4.46\%)} & \cellcolor{green!10}\textbf{0.0258(+7.05\%)} & \cellcolor{green!10}\textbf{0.0101(+12.22\%)} & \cellcolor{green!10}\textbf{0.0141(+9.30\%)} & \cellcolor{green!10}\underline{0.0316} & \cellcolor{green!10}\underline{0.0370} \\ \cdashline{2-8} & BIGRec & 0.0286 & 0.0337 & 0.0128 & 0.0169 & 0.0437 & 0.0498 \\ & +RW & 0.0282 & 0.0331 & 0.0136 & 0.0162 & 0.0422 & 0.0493 \\ & +SAM & \underline{0.0294} & \underline{0.0347} & 0.0135 & 0.0167 & \textbf{0.0446} & \textbf{0.0519} \\ & +GroupSAM & 0.0274 & 0.0316 & \underline{0.0149} & \underline{0.0177} & 0.0394 & 0.0449 \\ & \cellcolor{green!10}+EISAM & \cellcolor{green!10}\textbf{0.0303(+3.06\%)} & \cellcolor{green!10}\textbf{0.0358(+3.17\%)} & \cellcolor{green!10}\textbf{0.0160(+7.38\%)} & \cellcolor{green!10}\textbf{0.0203(+14.69\%)} & \cellcolor{green!10}\underline{0.0440} & \cellcolor{green!10}\underline{0.0506} \\ \midrule \multirow{10}{*}{\textbf{Steam}} & TALLRec & 0.0800 & 0.0889 & 0.0285 & 0.0420 & 0.1206 & 0.1259 \\ & +RW & 0.0811 & 0.0903 & 0.0291 & 0.0430 & 0.1221 & 0.1276 \\ & +SAM & 0.0824 & 0.0919 & 0.0283 & 0.0416 & \textbf{0.1250} & \underline{0.1315} \\ & +GroupSAM & \underline{0.0826} & \underline{0.0926} & \underline{0.0295} & \underline{0.0438} & 0.1244 & 0.1311 \\ & \cellcolor{green!10}+EISAM & \cellcolor{green!10}\textbf{0.0832(+0.73\%)} & \cellcolor{green!10}\textbf{0.0936(+1.08\%)} & \cellcolor{green!10}\textbf{0.0308(+4.41\%)} & \cellcolor{green!10}\textbf{0.0451(+2.97\%)} & \cellcolor{green!10}\underline{0.1245} & \cellcolor{green!10}\textbf{0.1318} \\
 \cdashline{2-8} & BIGRec & 0.0840 & 0.1018 & 0.0314 & 0.0462 & 0.1255 & 0.1456 \\ & +RW & 0.0872 & 0.1054 & 0.0391 & 0.0521 & 0.1251 & 0.1474 \\ & +SAM & \underline{0.0875} & \underline{0.1062} & 0.0388 & 0.0518 & \underline{0.1259} & \underline{0.1491} \\ & +GroupSAM & 0.0874 & 0.1059 & \underline{0.0395} & \underline{0.0531} & 0.1251 & 0.1475 \\ & \cellcolor{green!10}+EISAM & \cellcolor{green!10}\textbf{0.0887(+1.37\%)} & \cellcolor{green!10}\textbf{0.1098(+3.39\%)} & \cellcolor{green!10}\textbf{0.0411(+4.05\%)} & \cellcolor{green!10}\textbf{0.0564(+6.21\%)} & \cellcolor{green!10}\textbf{0.1262} & \cellcolor{green!10}\textbf{0.1519} \\ \midrule \multirow{10}{*}{\textbf{ADM}} & TALLRec & 0.0196 & 0.0231 & 0.0058 & 0.0088 & 0.0421 & 0.0464 \\ & +RW & 0.0185 & 0.0224 & 0.0063 & 0.0102 & 0.0384 & 0.0423 \\ & +SAM & \underline{0.0202} & \underline{0.0239} & 0.0050 & 0.0079 & \textbf{0.0449} & \textbf{0.0499} \\ & +GroupSAM & 0.0190 & 0.0227 & \underline{0.0069} & \underline{0.0107} & 0.0387 & 0.0422 \\ & \cellcolor{green!10}+EISAM & \cellcolor{green!10}\textbf{0.0216(+6.93\%)} & \cellcolor{green!10}\textbf{0.0258(+7.95\%)} & \cellcolor{green!10}\textbf{0.0079(+14.49\%)} & \cellcolor{green!10}\textbf{0.0116(+8.41\%)} & \cellcolor{green!10}\underline{0.0439} & \cellcolor{green!10}\underline{0.0489} \\ \cdashline{2-8} & BIGRec & 0.0205 & 0.0269 & 0.0075 & 0.0134 & 0.0417 & 0.0489 \\ & +RW & 0.0187 & 0.0243 & 0.0085 & 0.0148 & 0.0353 & 0.0398 \\ & +SAM & 0.0213 & 0.0281 & 0.0081 & 0.0142 & \underline{0.0428} & \underline{0.0507} \\ & +GroupSAM & \underline{0.0215} & \underline{0.0282} & \underline{0.0092} & \underline{0.0154} & 0.0415 & 0.0490 \\ & \cellcolor{green!10}+EISAM & \cellcolor{green!10}\textbf{0.0225(+4.65\%)} & \cellcolor{green!10}\textbf{0.0295(+4.61\%)} & \cellcolor{green!10}\textbf{0.0102(+10.87\%)} & \cellcolor{green!10}\textbf{0.0168(+9.09\%)} & \cellcolor{green!10}\textbf{0.0425} & \cellcolor{green!10}\textbf{0.0502} \\ \bottomrule \end{tabular}
% }
\label{tab:main}
\end{table*}

\subsection{Overall Performance (RQ1)}

\paragraph{Overall Performance.}
Table~\ref{tab:main} summarizes the overall recommendation performance of all methods across the three datasets. EISAM consistently surpasses the baselines on both $\mathrm{NDCG}@10$ and $\mathrm{HR}@10$ metrics. Specifically, EISAM achieves an average improvement of 3.53\% in overall $\mathrm{NDCG}@10$ and 4.54\% in $\mathrm{HR}@10$ compared to the strongest baseline. 

We observe that SAM consistently improves the overall performance compared with the original LRSs, while Reweight and GroupSAM do not necessarily yield positive gains and often introduce side effects. This is because SAM regularizes the model with respect to the global loss sharpness, thus enhancing the overall generalization capability. In contrast, Reweight and GroupSAM focu on local or group-specific adjustments without such a global guarantee, which may only bring localized optimization improvements. Our proposed EISAM, by introducing a fine-grained weighting mechanism combined with item-wise sharpness regularization, forms a new SAM variant that still optimies the global loss while providing more flexible item-level control. Consequently, EISAM achieves more stable and superior global performance than standard SAM. 

\paragraph{Tail Item Performance.}
We further evaluate the effectiveness of EISAM on tail items, which constitute 80\% of the item space. As shown in Table~\ref{tab:main}, EISAM achieves remarkable gains on tail items, with an average improvement of 8.90\% in $\mathrm{NDCG}@10$ and 8.44\% in $\mathrm{HR}@10$ over the best-performing baseline. 

Although SAM contributes to overall performance, it brings little improvement to tail items, indicating that the SAM framework itself still suffers from long-tail bias. GroupSAM provides relatively stable improvements on tail items, but its coarse-grained group-level control and the heavy computational cost of multiple SAM evaluations per group limit its effectiveness. Moreover, the lack of inter-group interaction is crucial for recommendation tasks, often results in unstable or even degraded overall performance. In contrast, EISAM introduces a novel item-wise sharpness that mitigates these issues, delivering consistent tail improvements with far lower overhead. These findings confirm that EISAM effectively alleviates the long-tail problem.

\begin{figure}[t]
  \centering

  \begin{subfigure}[t]{0.49\linewidth}
    \centering
    \includegraphics[width=\linewidth, trim=0cm 0cm 3cm 0.8cm, clip]{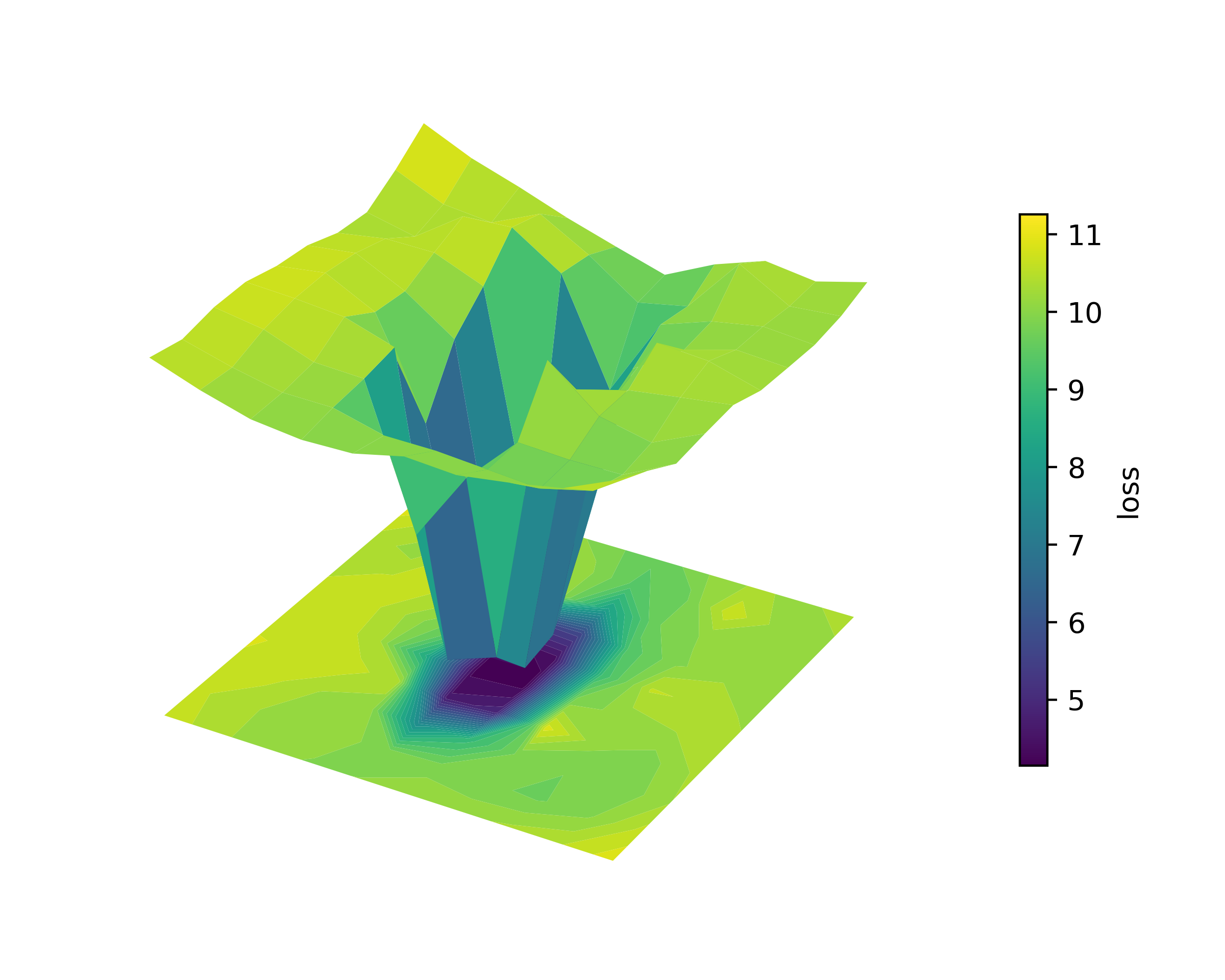}
    \caption{SAM on ML-1M}
  \end{subfigure}
  \begin{subfigure}[t]{0.49\linewidth}
    \centering
    \includegraphics[width=\linewidth, trim=0cm 0cm 3cm 0.8cm, clip]{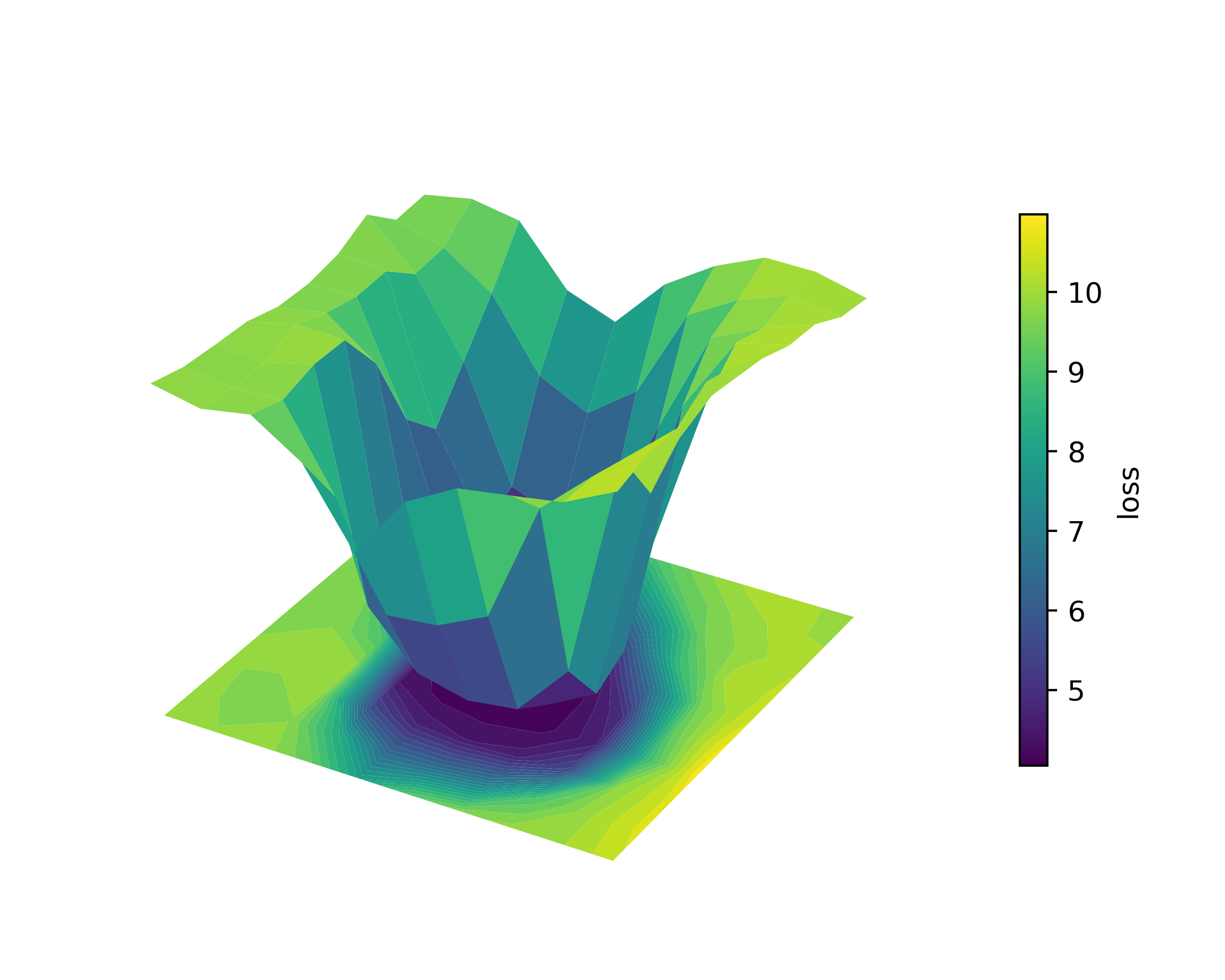}
    \caption{EISAM on ML-1M}
  \end{subfigure}
  \begin{subfigure}[t]{0.49\linewidth}
    \centering
    \includegraphics[width=\linewidth, trim=0cm 0cm 3cm 0.8cm, clip]{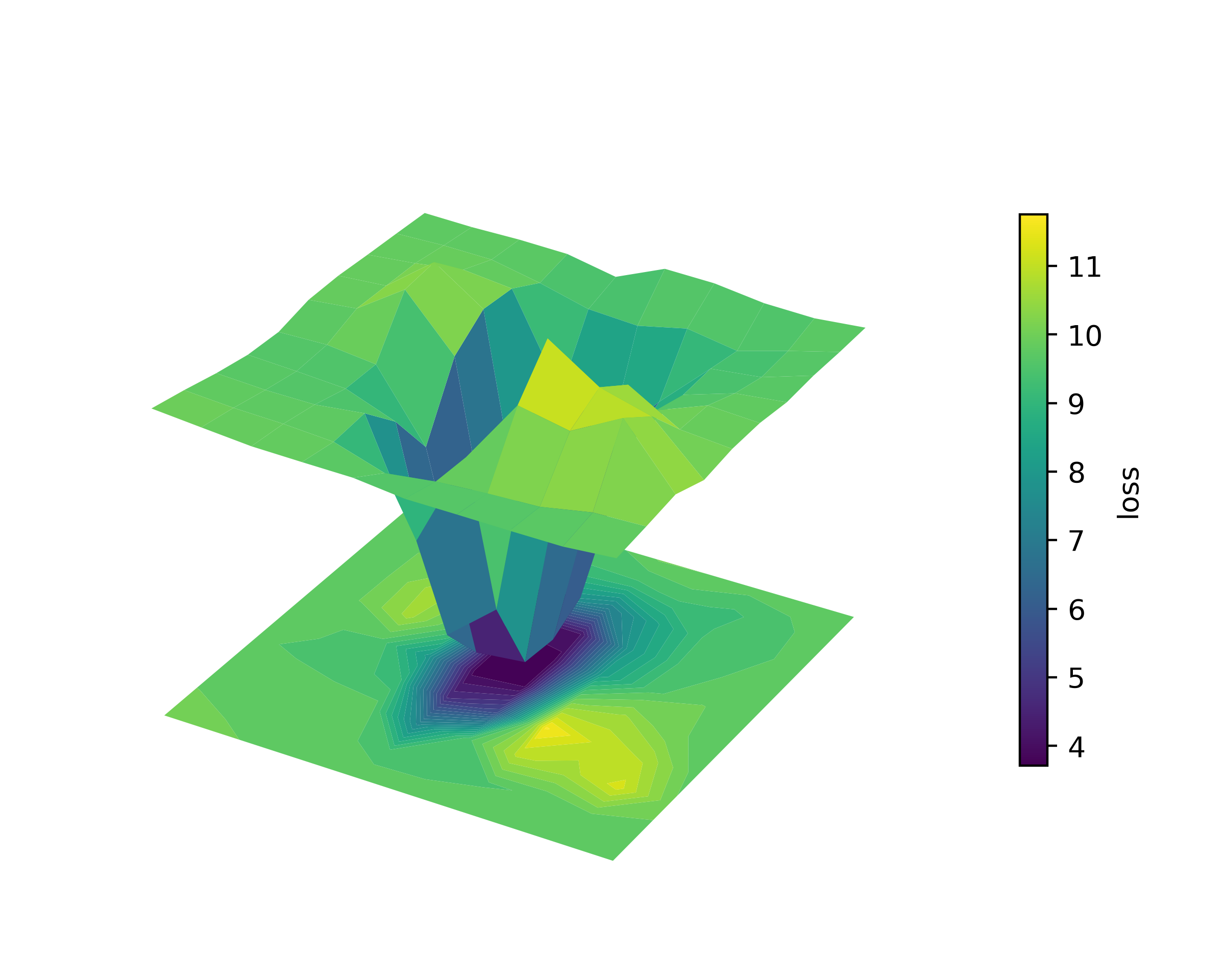}
    \caption{SAM on Steam}
  \end{subfigure}
  \begin{subfigure}[t]{0.49\linewidth}
    \centering
    \includegraphics[width=\linewidth, trim=0cm 0cm 3cm 0.8cm, clip]{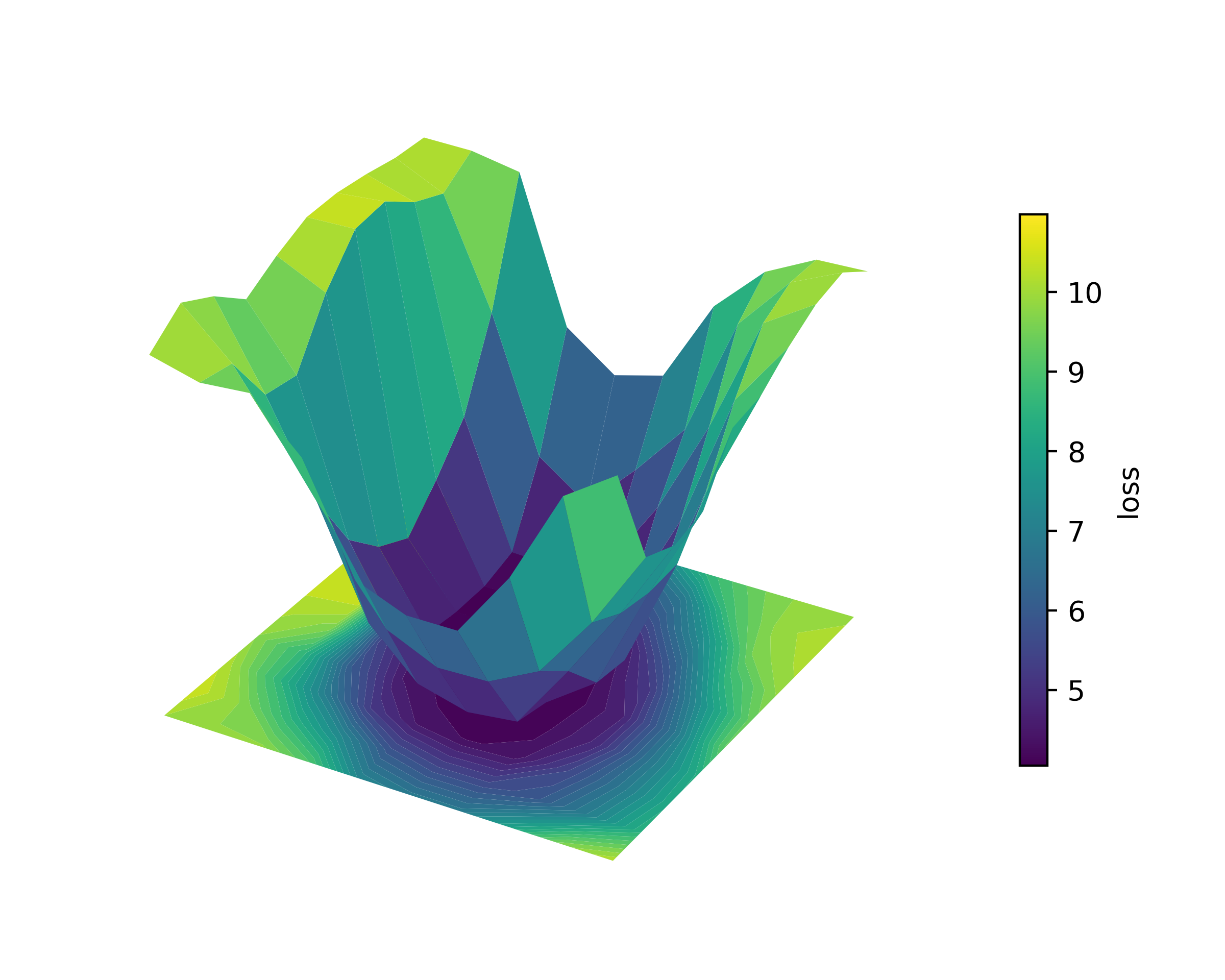}
    \caption{EISAM on Steam}
  \end{subfigure}
  \begin{subfigure}[t]{0.49\linewidth}
    \centering
    \includegraphics[width=\linewidth, trim=0cm 0cm 3cm 0.8cm, clip]{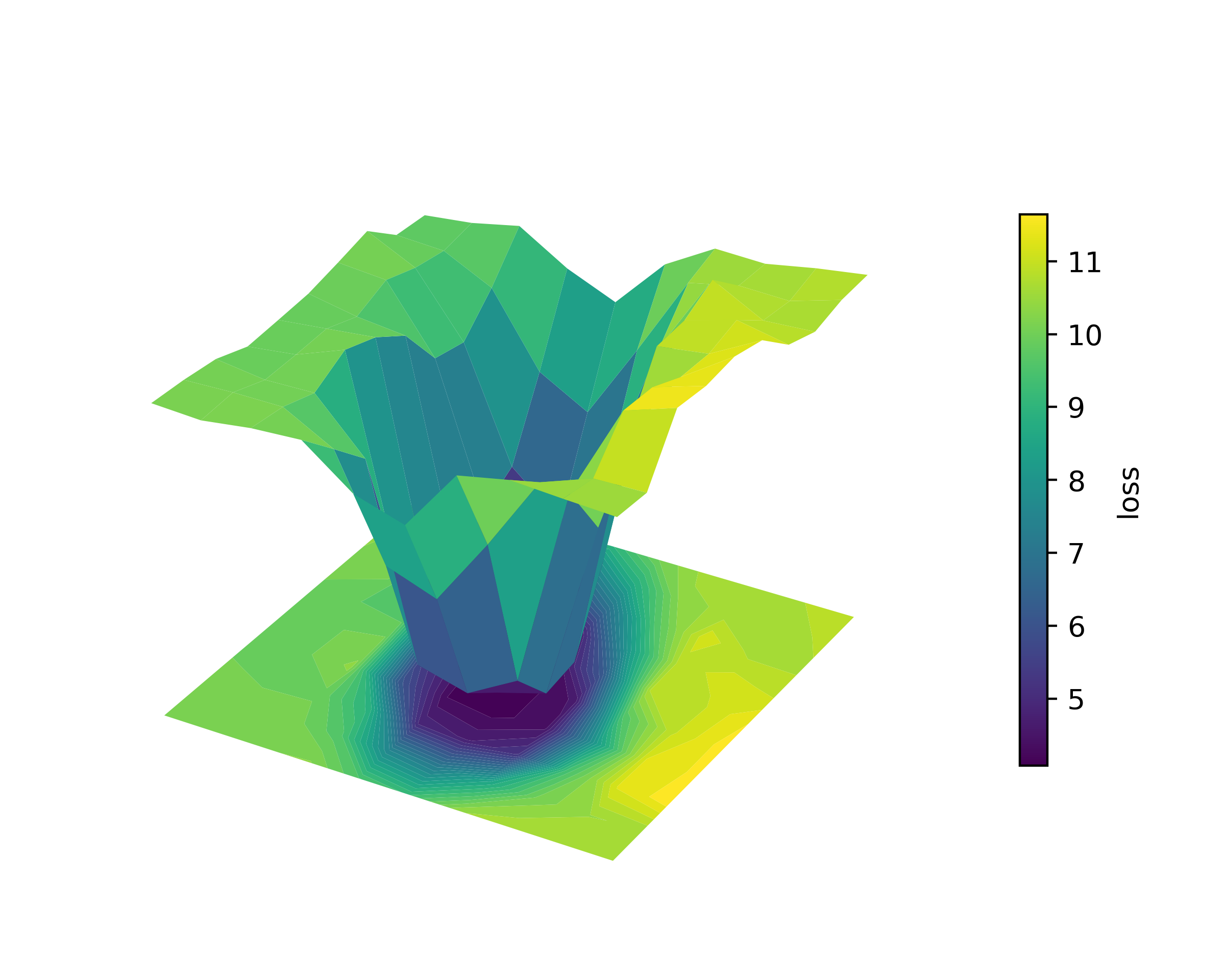}
    \caption{SAM on ADM}
  \end{subfigure}
  \begin{subfigure}[t]{0.49\linewidth}
    \centering
    \includegraphics[width=\linewidth, trim=0cm 0cm 3cm 0.8cm, clip]{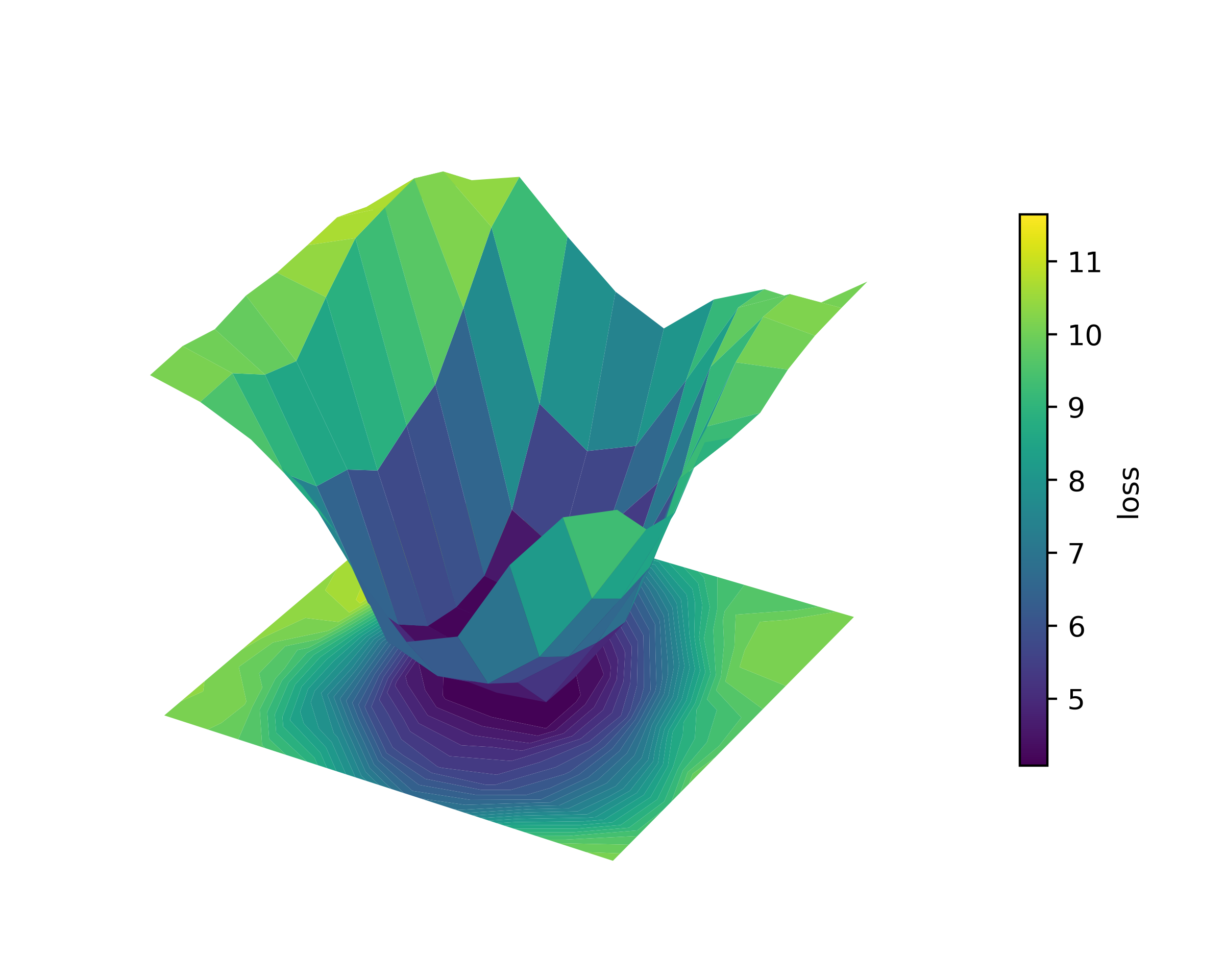}
    \caption{EISAM on ADM}
  \end{subfigure}
  \caption{visualizations of the loss landscapes of LRS on tail items. Darker colors indicate lower loss values.}
  \label{fig:landscape}
\end{figure}
\subsection{Loss Landscape (RQ2)}
To further investigate EISAM’s learning behavior on tail items, we visualize and compare the loss landscapes of EISAM and SAM in Figure~\ref{fig:landscape}. From both 3D and 2D perspectives, SAM exhibits limited flatness on tail items, indicating restricted generalization in long-tailed scenarios. In contrast, across all three datasets, EISAM demonstrates a markedly flatter and smoother loss surface, suggesting that its item-wise sharpness regularization effectively mitigates sharp minima and stabilizes optimization. This evidence highlights that, for tail items, EISAM better captures local curvature while preserving global smoothness, thereby improving generalization and robustness under long-tail distributions.
\subsection{Training Efficiency (RQ3)}
Table~\ref{tab:time} compares the computational efficiency among different methods. 
EISAM exhibits comparable training efficiency to standard SAM, introducing only an average 5.3\% overhead per epoch. On the Steam and ADM datasets, the additional computational cost is as low as 0.01\%.
EISAM achieves fine-grained item-wise sharpness regularization with only a small and acceptable number of additional computations, 
demonstrating its efficiency in modeling item-level curvature without heavy cost. 
This efficiency is attributed to our optimized training procedure, which avoids the group-level looping required by GroupSAM. 
In contrast, GroupSAM incurs approximately \textbf{Group×} higher time complexity. 
Therefore, EISAM achieves an optimal balance between accuracy and training cost, making it scalable to large real-world LRS applications.
\subsection{Impact of Hyperparameter (RQ4)}
\begin{figure}[t]
  \centering

  \begin{subfigure}[t]{0.49\linewidth}
    \centering
    \includegraphics[width=\linewidth]{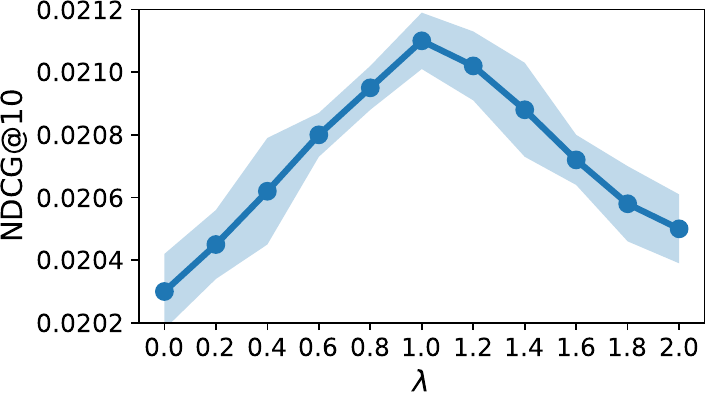}
    \caption{Impact of $\lambda$}
  \end{subfigure}
  \begin{subfigure}[t]{0.49\linewidth}
    \centering
    \includegraphics[width=\linewidth]{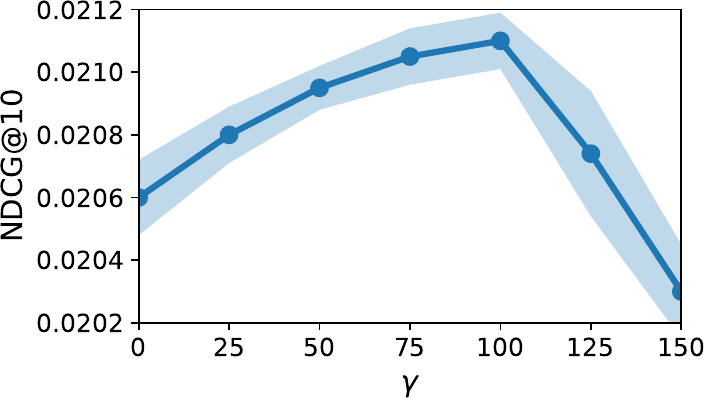}
    \caption{Impact of $\gamma$}
  \end{subfigure}

  \caption{Hyperparameter sensitivity analysis.}
  \label{fig:hyper}
\end{figure}

We analyze the sensitivity of EISAM to two key hyperparameters, $\lambda$ and $\gamma$. $\gamma$ serves as a weighting coefficient in the smoothly attenuated exponential weighting function $f_{\text{exp}}(\cdot)$ and adjusts the relative importance of tail items when computing item-wise sharpness. We conduct experiments on \textit{ML-1M} using \textit{BIGRec} as the base LRS and report the results over five independent runs. Figure~\ref{fig:hyper} illustrates their effects.

\begin{table}[t]
\centering
\caption{Average training time per epoch. Also show the relative factor in parentheses (w.r.t.\ SAM in the same column).}
\label{tab:time}
\resizebox{\columnwidth}{!}{%
\begin{tabular}{lccc}
\toprule
\textbf{Methods} & \textbf{ML-1M} & \textbf{Steam} & \textbf{ADM} \\
\midrule
SAM            & 3182s (1.00$\times$) & 3897s (1.00$\times$) & 949s (1.00$\times$) \\
GroupSAM       & 12172s (3.82$\times$) & 15844s (4.06$\times$) & 3948s (4.16$\times$) \\
\textbf{EISAM}   & 3652s (1.14$\times$) & 3969s (1.01$\times$) & 967s (1.01$\times$) \\
\bottomrule
\end{tabular}%
}
\end{table}

\paragraph{Impact of $\lambda$.}
We notice that increasing $\lambda$ initially improves model accuracy by promoting flatter minima but leads to degradation when $\lambda$ becomes overly large (around $\lambda>$~1.0). This reflects the trade-off between minimizing training loss and minimizing the sharpness of the loss landscape. Moderate $\lambda$ values achieve the best compromise, maintaining stability and robustness without excessive regularization. Overall, EISAM demonstrates strong robustness across a wide range of hyperparameter values, showing consistent improvements under the long-tail setting.

\paragraph{Impact of $\gamma$.}
As $\gamma$ increases, performance first improves, indicating that emphasizing the sharpness of tail items enhances generalization under long-tail distributions. However, when $\gamma$ exceeds a certain threshold (e.g., $\gamma >$~100), the performance begins to decline, suggesting that overemphasizing tail sharpness can lead to over-regularization and hinder learning on head items. This finding implies the necessity of balancing head and tail regularization strengths in the optimization process.
\section{Conclusion}
In this work, we systematically investigated the long-tail problem in LRSs, distinguishing between the \textit{prior long-tail} from pretraining and the \textit{data long-tail} in recommendation data. 
Our empirical analysis shows that both of them affect performance, but the data long-tail remains the dominant factor shaping head and tail disparities. The prior long-tail mainly enhances the performance of head classes, while exerting limited influence on the tail.
To mitigate this issue, we propose \textbf{Efficient Item-wise Sharpness-Aware Minimization (EISAM)}. 
EISAM adjusts the loss curvature per item, enabling finer control over head and tail groups. Moreover, we design an efficient optimization procedure to alleviate the high computational overhead, ensuring that EISAM remains both effective and practical for LRSs.
Experiments on real-world datasets show that our method improves overall performance, significantly enhances the recommendation quality for tail items, and maintains computational efficiency. Moreover, empirical evidence indicates that our approach successfully reduces the sharpness of tail items.
\begin{acks}
    This work was supported in part by the National Natural Science Foundation of China (No.~62522217, No.~62402148) and Ant Group.
\end{acks}

\bibliographystyle{ACM-Reference-Format}
% \balance
\bibliography{sample-base}

\balance

\end{document}